\documentclass[onecolumn, preprint, review, authoryear,a4paper,10pt,oneside]{elsarticle}
\usepackage{newtxtext,newtxmath}
\usepackage[applemac]{inputenc}
\usepackage[a4paper,left=3cm,right=3cm]{geometry}
\usepackage{natbib}
\usepackage{amsfonts}
\DeclareMathSymbol{\R}{\mathalpha}{AMSb}{"52}
\DeclareMathSymbol{\I}{\mathalpha}{AMSb}{"49}
\usepackage{mathrsfs}
\journal{arXiv}
\begin{document}
\begin{frontmatter}
\title{A neo-Kantian explanation of the applicability of mathematics to physics}
\author[jm]{Jorge Manero\corref{cor1}}
\ead{jorge.manero@filosoficas.unam.mx; ORCID: 0000000345885132}
\cortext[cor1]{Corresponding author}
\address[jm]{Universidad Nacional Autónoma de México, Instituto de Investigaciones Filosóficas,  Circuito Maestro Mario de la Cueva s/n, Ciudad Universitaria, 04510, Coyoacán, Mexico City, Mexico.}
\begin{abstract}
Various `optimistic' attempts have been made to reasonably explain the undeniable effectiveness of mathematics in its application to physics. They range over \emph{retrospective}, historical accounts of mathematical applicability based on pragmatic considerations, on the one side, and \emph{prospective} accounts based on indispensability considerations, on the other. In view of some objections that I will raise against these accounts, I would like to propose a third alternative based on Ernst Cassirer’s neo-Kantian view which can overcome these objections and embrace both pragmatic and indispensability considerations. According to this view, mathematics and physics are seen as different modes of a basic process of cognitive synthesis that are essentially applied to each other according to \emph{a priori} principles of theory development inherently incorporated into scientists’ minds. As emphasised by Cassirer, these principles have a constitutive role (i.e., they explain the relevant phenomena across scientific change) and a regulative role (i.e., they incorporate the ideal of unity, permanence and generality at each stage of knowledge), both of which have been captured through a functional (i.e., structural) understanding of concepts. As a particular case study, these principles shall be instantiated by invariance groups responsible for the effectiveness of applying Lie group theory to physics. 
\end{abstract}
\end{frontmatter}
{\bf keywords}: Cassirer's neo-Kantianism; Application of mathematics to physics; Invariance groups; Theory change; Surplus structure.
\section{Introduction}\label{introd}
In the introduction to his celebrated 1931 paper, Dirac asserts that 
\begin{quote}
[...] The most powerful method of advance that can be suggested at present is to employ all the resources of pure mathematics in attempts to perfect and generalise the mathematical formalism that forms the existing basis of theoretical physics, and after each success in this direction, to try to interpret the new mathematical features in terms of physical entities (by a process like Eddington’s Principle of Identification) \citep[60]{dirac}.
\end{quote}
As emphasised by \citep[11]{ryckman}, what underlies this powerful assertion is the method of mathematical conjecture in fundamental physical theory. It is undeniable that this method is effective as involving a claim about the expressive, innovative and interpretative power of mathematics and its effective application to physical science. However, there are plenty of questions that arise from taking this and other effective methods at face value. First and foremost, why is this method so effective? What does ‘advance’ mean in the above quotation? With respect to which aspects is it effective? Is it a pragmatic method valid for certain physical theories or is it an infallible method of theory construction? 

Thus far there is no single, authoritarian dogma that can answer all these questions at once, but a plurality of views addressing some of the most relevant issues, at best. In particular, explanations of the effective application of this and other methods of theory construction range over \emph{retrospective}, historical accounts of mathematical applicability based on pragmatic considerations, on the one side, and \emph{prospective} accounts based on indispensability considerations, on the other. Whereas the former do not focus on the necessary and indispensable roles appropriate to the constitutive dimension of mathematical knowledge, the latter ignore the pragmatic and normative roles, appropriate to its regulative dimension. However, although these explanations might seem opposites in certain essential respects, both presuppose a reifying dualism between the mathematical and the physical compartments. That is, according to both explanations, the above method supposedly establishes a set of correspondence rules between the abstract reign of mathematical conjecture and the concrete reign of physical theorising, the former being seen as an uninterpreted calculus and the latter as the interpretative arena under which empirical knowledge is systematised and conceptualised. 

Considering that an appropriate, complete explanation of the effective application of this and other methods has not been established, my plan is to undermine the reified dualism between the mathematical and the physical realms and advocate a contemporary revival of the Kantian method of \emph{a priori} mathematical conjecture in fundamental physical theory, in which both the `intellectual’ and the `intuition’ compartments of cognition are deeply intertwined. Thus, instead of articulating a representational framework in terms of which mathematics and physics are externally related via certain applicability constraints or correspondence rules, I wish to explore the underlying transcendental conditions for the mere possibility of bringing empirical and physical knowledge under mathematical concepts (and vice versa) in the first place. The conditions to be unveiled in this Kantian analysis are, to put it crudely, the transcendental presuppositions (i.e., the lens through which all humans observe) by means of which the mind perceives, conceptualises, and mathematises the phenomena (i.e., the objects that affects our senses). 

In so doing, I shall rely on the \citep{heis2014a,ryckman} reading of Cassirer’s neo-Kantianism relevant to the issue of mathematical applicability, which put forward the view that the above conditions are established by a set of \emph{a priori} principles of theory development inherently incorporated into scientists’ minds, as they are conditions for the possibility of knowledge that escape any form of sensation and definite conceptual construction. However, \emph{contra} the fixed Kantian \emph{schema}, these principles change and evolve in a progressive manner such that they can only be partially instantiated at each stage of knowledge. As emphasised by Cassirer, given the development of non-Euclidean geometry and its application to general relativity, these principles cannot be associated with an \emph{a priori} conception of spacetime, as Kant himself thought, but with regulative, normative ideals (i.e., unity, permanence and generality), and constitutive, epistemic virtues (i.e., predictive and explanatory power), in terms of which the diversity of experience is explained \citep{ryckman,heis2014a}. In accordance with these ideals and virtues, Cassirer interprets the alleged principles of theory development from the standpoint of the \emph{function theory of concepts}, according to which the category of structures and relations standing between individual concepts is taken as primitive and more fundamental than the individual concepts themselves \citep{heis2014b}. The structural interpretation of these principles enables us to elucidate the constitutive, indispensable role of mathematical conjecture in its application to physics by virtue of explaining the relevant phenomena across scientific change, and also enables us to reveal the regulative role associated with certain pragmatic considerations which underlie the ideals of unity, permanence, and generality at each stage of scientific knowledge \citep{ihmig}.

Moreover, based on the presupposition that Lie group theory faithfully captures the structural interpretation of concepts, physicists and philosophers of neo-Kantian spirit have certainly combined efforts to represent the alleged \emph{a priori} principles in terms of this mathematical language via the identification of \emph{invariance groups}, such as \emph{general covariance} \citep{cassirer1923,ryckman}.\footnote{The invariance group associated with general covariance is the diffeomorphism group of fourth-dimensional spacetime. I shall define invariance group in section \ref{section3}.} However, they have not paid sufficient attention to the role played by invariance groups as laying down the \emph{a priori} principles in terms of which the effectiveness of Lie group theory can be explained. Under these circumstances, the last part of this paper intends to bridge this gap by characterising the group-theoretic properties of invariance groups responsible for the effective application of Lie group theory to physics.

My methodological plan shall be as follows. In Section \ref{section2}, I briefly sketch the problem of the applicability of mathematics to physics. In Section \ref{section3}, I present the neo-Kantian framework of mathematical applicability. In so doing, I shall write a brief introduction to Kantian epistemology and Cassirer’s neo-Kantianism \citep{cassirer1922,cassirer1923,cassirer1956}, both of which shall be discussed according to the \citep{ryckman,heis2014a} reading. It is important to note that I shall not interpret Kantian interpretations; I shall take some considerable time to explain this reading of Cassirer as it is essential to my proposal and there is not such a thing as `the canonical view' of the field. In Section \ref{section4}, I shall demonstrate the relevance of this novel framework applied to the domain of Lie group theory. Finally, in Section \ref{section5} I shall write some concluding remarks. 
\section{The problem of the applicability of mathematics to physical science}\label{section2}
The problem of the applicability of mathematics to physical science can be stated as the capability to investigate, characterise and explain the `logic’ behind the effective application of many mathematical structures beyond the mathematical and/or scientific domains within and/or for which they were originally developed. This problem has been the focus of ample and controversial discussion since the time of the ancient Greeks. However, in recent decades this problem has attracted considerable attention from prominent scientists and philosophers. 

At one end of the spectrum one finds \emph{mathematical pessimism} principally advocated by \citep{steiner} and inspired by Eugene Wigner’s claim that the ‘miraculous’ effectiveness of mathematical structures in their application to physical science cannot reasonably be explained, and that the extent and means by which these structures are successfully applied to this domain is ``something bordering on the mysterious” \citep[2]{wigner}. At the other end of the spectrum, however, one finds \emph{mathematical optimism}, according to which the prevailing effectiveness of mathematics in its application to physical science can actually be explained and characterised appropriately.\footnote{The distinction between mathematical pessimism/optimism was originally coined by \citep{wilson} and later adopted by \citep{buenofrench}.} 

The way philosophers advocate the optimistic end of the spectrum has varied according to their epistemological and/or metaphysical commitments with respect to mathematics and physics.\footnote{According to \citep{bangu}, there are at least four optimistic solutions: the `many failures' solution, the `fudging' solution, the statistical solution and the `empirical origins' solution. Although this way of characterising the applicability problem of mathematics is reasonable, I shall appeal to other distinguishing criteria aligned to epistemic and metaphysical considerations.} At first sight, one may suppose that the way these fields of knowledge are applied to each other rely on certain pragmatic considerations associated with their own aims and ideals, fully respecting their relative autonomy. From this point of view, mathematics serves as a toolbox for the purposes of physics (e.g., prediction, representation, explanation, unification, etc.), in the same way as physics serves as a toolbox for the central purposes of mathematics (e.g., consistency, generality, systematicity, etc.). However, the problem with pragmatic considerations of this sort is that the aforementioned toolboxes lack any indispensable status, in the sense that we cannot establish with these considerations alone a necessary correspondence which extrapolates the scheme of effective application from particular and contingent instantiations that have happened in the history of science to those that will happen in its future development. Let us examine this in more detail. 

Overall, we can differentiate between the pragmatic task of explaining such effectiveness in a retrospective manner, in the sense of unveiling the principles behind the different \emph{pragmatic} roles (i.e., predictive, representational, explanatory and unifying roles) played by mathematics in developing already known successful theories from the task of arguing (in a prospective manner) for its \emph{indispensable} role in developing new effective theories on the basis of infallible principles or recipes for theory construction. For example, through a selective and case-by-case historical analysis of concrete case studies, adherents to the retrospective approach to mathematical optimism, such as \citep{grattan,wilson,buenofrench}, have attempted to make formal (or qualitative) comparisons between the physical and mathematical domains, thereby setting (or looking) out a series of ‘special physical circumstances’ by which the latter has been applied to the former (and the moves made on both sides) in close connection with actual scientific practices. Other influential advocates of this retrospective approach are \citep{leng,pincock}. Alternatively, adherents to the prospective approach to mathematical optimism, such as \citep{colyvan,tegmark,bangu} and other advocates of the well-known Quine-Putnam indispensability thesis (or its variants), have advanced different realist views about mathematics (e.g., neo-platonism, `Pythagorean' structuralism, explanationist realism) to elucidate an ontological explanation of this effectiveness based on the claim that mathematics plays an indispensable role. 

Thus far many attempts have been made to support or criticise these optimistic views on the applicability of mathematics. Without getting embroiled in details, I shall concentrate on two critical comments of my own that bear upon these views. 

With regard to the retrospective-pragmatic view, I see that a faithful, case-by-case reconstruction of the history of mathematical applicability is central to this optimist account. However, there is a strong whiff of contingency in any case-by-case historical analysis of this sort because it conceives history as a series of `snapshots’ without supplying an overall, continuous and systematic explanation of mathematical effectiveness that does not depend so much on the particular features or the `special circumstances’ of the historical moves involved. Furthermore, without stressing the indispensability role played by the mathematical linguistic framework in achieving any form of innovation in any of the many other scientific contexts (e.g., via mathematical conjectures or analogies), a `looking-forward’ view that provides a more general explanation embracing broader domains of applicability is excluded from the start. 

With regard to the prospective view, however, neo-Platonism, `Pythagorean' structuralism, explanationist realism (or alternative realist views) lack any connection with the realm of actual practice and cannot establish alone the special circumstances in which mathematics is scientifically relevant. As evidenced by actual scientific practices, there is no single, authoritarian dogma of mathematical applicability, but a series of pragmatic, context-dependent applications, at best. Indeed, any productive relation between the mathematical and scientific domains relies on the presupposition that there must be a \emph{context} and a \emph{purpose} in terms of which one of them is pressed into service to be effectively applied to the other. For example, mathematics can take a predictive, representational, explanatory or unifying role, and the degree of effectiveness with respect to each of these roles varies depending on the domain of its application. 

In view of these objections, I would like to propose a third alternative which can resolve the above problems and embrace both pragmatic and indispensability considerations. To do so, I shall identify a presupposition deeply involved in both the retrospective and prospective views, which will be shown to be responsible for these problems, namely, the distinction between the \emph{form} and the \emph{content} of the scientific object. For mathematical optimists (either in the form of neo-platonists, structuralists, explanationists, pragmatists or empiricists), the form of the scientific object is given by the purely mathematical language or conceptual representation, i.e., the mental contribution to the knowledge of this object, whereas the content is given by the scientific or empirical interpretation of this formalism, i.e., the object in itself. As we shall see, this distinction has profound consequences for the extent to which mathematics and physics are related and interpreted, for if they are found to belong to independent categories of knowledge, the way they apply to each other can be obscured by a mist of mystery that disassociates the mental, representational form of the object from its objective content. 

Following a particular reading of \citep{ryckman,heis2014a} about Cassirer’s philosophy, my contention is that an epistemological neo-Kantian view can reasonably explain the effectiveness of mathematics in its application to physics via a prospective approach that elucidates the merely objective, \emph{a priori} conditions for the possibility of scientific knowledge without excluding significant pragmatic considerations. It is via the identification of these conditions or principles that we are able to explain something which has long appeared a mystery, and for which extreme solutions, such as neo-Platonism or `Pythagorean' structuralism, have been seriously proposed. As we shall see, what needs to be sacrificed to advocate this Kantian alternative is the reification of the distinction between the form and the content of the scientific object. With this idea in mind, let us now try to unveil the appropriate Kantian framework. 
\section{A neo-Kantian framework of mathematical optimism.}\label{section3}
\subsection{Kant on the possibility of knowledge}
The central question associated with Kant’s transcendental idealism is purely epistemological, deeply motivated by a strong desire to understand science, human reason and the intricate nature of the mind.\footnote{The primary and secondary sources of this overview on Kantian epistemology are included in the Pedro Rivas translation of the \emph{Critique of Pure Reason} \citep{rivas}.} In a nutshell, the question at issue is the critical revision of the transcendental constitution of objectivity (i.e., beyond that revealed by the senses) via the systematic analysis of the conditions for the possibility and extension of \emph{synthetic a priori} knowledge.\footnote{As is well known, judgements (or concepts) of this kind are \emph{a priori} by virtue of being necessary and universal propositions absolutely independent of experience, whereas they are synthetic by virtue of being constituted by predicates not contained in the concept of the subject (e.g., mathematics and physics).} As already mentioned in the introduction, the conditions to be unveiled in the Kantian analysis are, to put it crudely, the transcendental presuppositions (i.e., the lens through which all humans observe) by means of which the mind perceives and conceptualises the phenomena (i.e., the objects of sensibility). In this way, Kant conceives nature not as an external element independent of the mind but as an object of knowledge, i.e., as a phenomenon which is unthinkable apart from the coexistent subjects capable of experiencing and conceptualising that object. 

On the basis of this essential question, Kant develops a systematic categorisation of the synthetic \emph{a priori}  according to many distinctions, but I shall only focus on one of the most relevant of them. Firstly, it is claimed that one can only obtain substantial, empirical knowledge of the world via our \emph{sensibility} (i.e., the human capacity of sense experience) and \emph{understanding} (i.e., the capacity to form thoughts about such experiences and to consider their conceptual formation). These are regarded as independent (though mutually interacting) sources of knowledge that cooperate via a mediating third, transcendental \emph{schema} or rules (both intellectual and sensible) for forming empirical judgements. The transcendental \emph{schema} consists in \emph{a priori}, inherently and universally valid \emph{pure intuitions} (in the case of its sensible dimension) or \emph{categories} (in the case of its conceptual dimension), such that there exists a single univocal cooperation between the realm of the understanding and that of the sensibility.\footnote{Note that contrary to any form of empiricism or metaphysical realism (sometimes, but not always associated with representationalism), the understanding-sensibility distinction is not reified as implying a correspondence between our conceptual representations and the empirical world; both poles of the distinction are rather regarded as different moments or modes of a basic process of cognitive synthesis solely distinguished for the sake of analysis. This means that the phenomena that we sense and the conceptual edifice that we articulate from such phenomena cannot escape the realm of ideas and are inherently interrelated ways to know not the world itself but the way we know the world via our sense capabilities, on the one hand, and our intellectual approximations, on the other.} 

Secondly, based on a sharp distinction between understanding (i.e., the constitutive form of knowledge) and \emph{pure reason} (i.e., the regulative), Kant engages in the ‘propaedeutic’ task of investigating the limits and boundaries of pure reason’s willingness to offer knowledge of the transcendental world. According to Kant, pure reason is associated with the regulative, necessary ideal of unity in its more general conception, excluding from its faculties any possibility of synthetic \emph{a priori} knowledge. This means that whereas reason seeks `absolute unity’ by virtue of the ultimate coherent use of concepts within the realm of the understanding, sensibility and understanding seek `relative unity’ by virtue of the coherent articulation of the diversity of experience. In particular, pure reason ascertains whether a judgement is true on the basis of how far it connects with other judgements of the understanding in the ideal end point of inquiry. However, whilst it plays an indispensable role in experience, it does not play a constitutive role (i.e., it is not necessarily realised in experience), as opposed to the principles of the understanding, which are the sources of the synthetic \emph{a priori} and a constitutive factor as regards the possibility of experience. 

With this brief overview, let us now return to our main concern. Central to Kant’s analysis of the constitution of objectivity is the fundamental status of mathematics and science, both of which are regarded as realised possibilities of synthetic \emph{a priori} knowledge that unveil the conceptual bridge between the understanding and sensibility. In this regard, Kant argues (as opposed to what later become the school of logical positivism) that whereas synthetic \emph{a priori} judgements, such as mathematical statements, are true by virtue of their application to experience, the possibility of experience is based on the synthetic conceptual unity of the phenomena, without which there can be no mathematical knowledge at all. So, rather than viewing mathematics purely as a system of implicit definitions devoid of physical content that is contingently connected to experience (i.e., as the form of knowledge), he understood mathematics as essentially and necessarily applicable to physics. In this way, the effectiveness of mathematics in its application to physics can be explained by the fact that both domains are necessarily tied together by a cognitive process of synthesis, in the sense that there can be no relevant move on the mathematical side without such a move implying a correspondence with the physical and empirical domains. 
 
However, although some of the central tenets of Kant’s epistemology can still be valid in the light of current knowledge, the majority of the arguments concerned with mathematics and science elaborated by him do not hold these days, such as the well-known assumption that Euclidean space and absolute time are the \emph{a priori} arena or form of intuition in which phenomena are organised in the mind, and the association of science with Newtonian physics, a theory which is considered today a restricted, idealised model of the directly observable world. 

Under these circumstances, Cassirer’s neo-Kantian approach can be considered a revival of transcendental idealism in the light of new discoveries in fundamental physics. As we shall see, contrary to other forms of neo-Kantianism (e.g., Hans Reichenbach’s philosophy \citep{reich}, or more recently Michael Friedman’s interpretation), the \citep{ryckman,heis2014a} reading of Cassirer tries to preserve Kant’s refusal to reify the understanding-sensibility distinction demarcating itself from any form of empiricism and metaphysical realism. However, \emph{contra} Kant, this neo-Kantian revival only makes sense if the \emph{a priori} principles that warrant any kind of scientific inquiry are not immune to continuous change and are relativised to a given stage of knowledge dissolving with it the constitutive and regulative distinction. 
\subsection{Cassirer’s neo-Kantianism}
According to \citep{ryckman,heis2014a,everett2,everett}, Cassirer revives Kant’s transcendental idealism mixed with a flavour of naturalism, according to which scientific knowledge is the most secure and genuine form of knowledge. Essentially, he investigated the conditions for the possibility of scientific knowledge via the identification of \emph{a priori} principles within accepted theories in the physical sciences. In so doing, he relied on three central premises: 
\begin{enumerate}
\item[(i)] As an advocate of the Marburg tradition of neo-Kantianism, Cassirer rejected the positivists’ reification of the distinction between the mental faculties of sensibility and understanding (which had been interpreted by Moritz Schlick as the `content’ and `form’ of knowledge, respectively). \\
\item[(ii)] Apart from obvious contextual differences, a central disagreement with Kant is that Cassirer rejected the distinction between understanding (i.e., the constitutive) and pure reason (i.e., the regulative), the latter of which expresses the ideals of unity, permanence and generality.\footnote{NB: as argued in \citep{cassirer1923} and emphasised by \citep{heis2014a}, Cassirer's target was not the idea that, for certain principles, it might help to draw the constitutive-regulative distinction; rather his target was the absolute dichotomy between constitutive and regulative principles.} \\
\item[(iii)] Although Cassirer retained the constitutive standing of \emph{a priori} principles relativised to a stage of scientific knowledge, he denied \emph{contra} Kant that such principles are necessarily valid for all times, and constitutive factors as regards the possibility of experience. As what later became Reichenbach’s notion of the \emph{relative a priori} developed in \citep{reich}, he asserted that principles are constitutive in the sense of delimiting the space of possible objects, but nonetheless not immune from experience, changing with the advance of physical science. 
\end{enumerate}
However, as \citep{ryckman,heis2014a,everett2,everett} acknowledge, in conjunction these premises are problematic, firstly owing to the problem of the constitution of scientific objectivity in the face of theory change; and secondly, due to the problematic question of how (physical) phenomena can be brought under (mathematical) concepts (and vice versa) without any fixed \emph{schema} whose sensibility dimension is given by pure intuition i.e., space and time in their original conception. Let us address each problem in turn. 
\begin{enumerate}
\item[(I)] \emph{The problem of objectivity}. As emphasised by \citep[111]{friedman} and \citep{friedman2}, pure reason in its original Kantian conception consists in regulative, non-constitutive principles expressed in the form of ultimate ideals of total unity and permanence, upon which the notion of \emph{objectivity in general} is based. This means that these principles only reflect, at every step of the development of science, the tendency to search for permanent and invariant connections over and above the changing features of experience i.e., they ensure that there is a logical structure in the sequence of theories as they are investigating the same subject. \emph{Contra} the \citep{ryckman,heis2014a} reading of Cassirer, the ideals of total unity and permanence can provide, at best, a retrospective explanation of the rationality of adopting new theories with respect to an ideal limit, whereas they cannot constitute current scientific knowledge. The reason, according to Friedman, is that regulative principles of this sort are insufficient to enable the empirical and physical interpretative side of theories. In this way, the \emph{a priori} principles of pure reason that constitute all experience are never known in advance and will only be known at the end point of scientific inquiry via \emph{supreme laws} or \emph{norms} that guide the formation of all physical theories in the history of human thought. 

In contrast to Friedman's interpretation, \citep{ryckman,heis2014a} claim that Cassirer not only considers the role of supreme laws of norms in accounting for objectivity in general, but also expands the Kantian limits of pure reason to the realm of sensibility by relativising the constitutive role of the regulative ideal of unity and permanence to each stage of scientific development. In agreement with the above premise (ii), they argue that Cassirer’s methodology identifies the objective, \emph{a priori} principles of science that not only express the regulative aims and ideals of scientific knowledge at each stage of development, but also constitute the objects of science confined to certain relevant domains, as \citep{reich} emphasises in the notion of the relative \emph{a priori}. Whereas he interprets these principles as constitutive \emph{a priori} —as they are necessarily universal elements of the understanding-sensibility process of synthesis— these principles are considered partially objective (i.e., an instance of what is often called \emph{objectivity at-a-time}) by virtue of being associated with the relativised, regulative ideal of unity, permanence and generality, in terms of which the diversity of experience is structured at each stage of scientific knowledge. In this sense, objectivity in general, as opposed to objectivity at-a-time, is not given but is the ultimate goal at which to arrive through successive stages of the concept of object.\footnote{This has been expressed by \citep{ryckman} in terms of a step-by-step process of `deanthropomorphisation’ going from preceding, discarded theories to more robust and general theories, the last which contain fewer traces of subjective content.}

However, as criticised by \citep{friedman2010b,friedman2010a}, the problem with this reading of Cassirer is that under the partial notion of objectivity at-a-time, although the regulative-constitutive distinction is dissolved, there cannot be constitutive invariants under scientific changes, undermining the above premise (iii). In particular, objectivity at-a-time ``sheds no transcendental light on the actual historical process by which we arrived at general relativity in the first place'' \citep[783]{friedman2010a}. This is the reason why Friedman follows Reichenbach’s notion of the relative \emph{a priori} in a different way to Cassirer’s method as he intends to provide the conditions for the possibility of a given scientific theory instead of providing those of the entire historical sequence of physical theories \citep{everett}. Furthermore, as emphasised by \citep[186]{friedman2010b}, there is no way to provide a prospective account of how new conceptual content has emerged from the old. Under this `forward-looking’ account, one needs not only to identify the logical, invariant structures of theory development that persist from theory to theory and are common to all the possible most general forms of experience among the entire sequence of theories, but also to establish the rational ground behind the abandonment of an established conceptual framework in favour of a new one.\\
\item[(II)] \emph{The problem of discontinuous schema}. Let us recapitulate that intuition is, according to Kant, the family of representations through which knowledge immediately refers to phenomena. Whereas \emph{empirical intuitions} are produced by the effect of an object on the human senses, the diversity of sensations is articulated and ordered through \emph{a priori} representations known as \emph{pure intuitions}, which are the unsensed ground of phenomena which allows us to acquire empirical knowledge and bring sensations under a single unity of relations within the realm of sensibility. Thus, if we exclude the concepts of an object from its representation, going from the realm of the understanding to that of sensibility (e.g., substance, divisibility, etc.), and if we also extract any form of sensibility from this representation (e.g., colour, impenetrability, etc.), we end up with a pure form of sensibility devoid of any sensed content, which according to Kant is given by \emph{space} and \emph{time}. In this way, space and time are the unsensed ground \emph{schema} in terms of which sensible data is brought under a systematic unity (and vice versa) at the level of the physical phenomena. 

Provided that the space and time structures of contemporary theories are not the same as those of the science developed in Kant’s era, many attempts have been made to replace the notion of pure intuition by other space and time structures that are compatible with contemporary science. Well-known examples are Minkowski spacetime in special relativity or the dynamical spacetime structure of general relativity (something which is problematic due to the field-theoretic nature of spacetime in this theory). However, we cannot fix the \emph{schema} that corresponds to the space and time structures of a particular theory without problematic consequences, given that such structures are not immune to future changes. We are definitely confronted with a risk of arriving at a similar situation to that of the Newtonian-relativistic transition in which the space and time structures abruptly changed. Under these circumstances, another constitutive problem that Cassirer’s neo-Kantianism faces is how to bring (physical) phenomena under (mathematical) concepts (and vice versa) without the introduction of any fixed, discontinuous \emph{schema} \citep{everett2}. 
\end{enumerate}
In order to respond to both problems, Cassirer addressed a missing part of Kant’s critical analysis: the problem of the transformation and constitution of the (empirical and mathematical) concept in physical theory. As a consequence of the development of new physical theories (e.g., relativity theory, quantum mechanics, etc.), Cassirer argued that the mere constitution of the scientific concept shifted from the \emph{substance} to the \emph{function theory of concepts}, the latter which is an epistemological stance expressed by the mathematical notion of `function’ and encapsulated by the transcendental object-constituting relation of `intellectual coordination' \citep{ihmig,heis2014b}. As discussed in \citep{everett2}, two different aspects underlie Cassirer’s distinction between substance and function concepts: firstly, the type of concepts that should be used in science; and secondly, the kind of epistemological theory that should be adopted in relation to these concepts. 

With regard to the first aspect, the substance theory interprets concepts as self-subsisting individuals e.g., objects as bearers of a bundle of intrinsic properties, whereas the function theory put forward the view that the category of structures and relations (e.g., mathematical equations) standing between individual concepts are taken as more (epistemically) fundamental than the individual concepts themselves. 

With regard to the second aspect, the substance theory of the scientific concept draws a sharp distinction between the purely mathematical dimension (i.e., the form) and the physical or empirical dimension of the concept of science (i.e., its content), thus excluding with this distinction the epistemic (Kantian) preconditions of various kinds of knowledge. This is because behind this theory lie two epistemic and metaphysical presuppositions that undermine any form of \emph{a priori} scientific knowledge: \emph{metaphysical (scientific) realism}, according to which scientific concepts provide an approximately true, objective description of the behaviour of external objects as they are assumed to live in the external world independently of the mind; and \emph{epistemological atomism}, which claims that conceptual and empirical knowledge can be acquired directly by itself, rejecting with this claim any mathematical or theory-laden assumption that shapes the objects as they are conceptualised or observed \citep[267]{cassirer1922}. In contrast, the epistemological characterisation of the concept of physical theory in terms of the function theory blurs the distinction between the purely mathematical and the physical dimensions of the scientific concept. This means that, \emph{contra} metaphysical realism, concepts such as `truth’, `knowledge’ or `object’ are all explained in terms of `objectivity’ (and not the other way around), whereas \emph{contra} epistemological atomism, no scientific concept can be acquired all on its own and no single statement can be confirmed in isolation \citep[391]{heis2014a,cassirer1923}. 

Adopting this structural stance that echoes Duhem’s holism, Cassirer is able to respond to the first problem of objectivity and scientific change by elucidating the notion that there cannot be knowledge of an object outside a total system of structural concepts and \emph{a priori} principles that is (and will be) preserved as science develops. The preservation of this system across theory change allows us to justify the abandoning of one stopping-point for another because, over and above the unexpected variation in the natures of the objects of science, the entire sequence of physical theories complies with the regulative, constitutive demands of unity, permanence, and certain principles of theory selection (e.g., mathematical simplicity and generality), all of which reinforce the rationality of theory change. These concepts and principles are constitutive because they explain the relevant phenomena at each stage of scientific development, whereas they are regulative by virtue of maintaining ``[...] a general logical structure in the entire sequence of special conceptual systems.'' \citep[16]{cassirer1922}. Analogously, Cassirer is able to neutralise the second problem by stipulating that the objects of scientific theory, including both theoretical and observable entities, are mathematically-laden, in the sense that these objects cannot be known and conceptualised outside a system of structural, mathematical concepts and \emph{a priori} principles that indispensably explain the physical phenomena \citep{heis2011,heis2014a}. 

Returning to the relationship between mathematics and physical science, the possibility of incorporating both the constitutive and regulative elements into the way scientific knowledge is developed at every successive step sheds some light upon the reciprocal role played by these domains. Indeed, mathematics and physics stand on each other’s shoulders by virtue of the regulative and constitutive ideal of pure reason uniting every aspect of the understanding and sensibility, such as the conceptualisation and generalisation of natural regularities in terms of mathematical structures. This could be the corroboration of the claim that mathematics and physical science are applied to each other according to certain pre-established aims, some of which are the regulative ideals of permanence and generality. In this way, it makes sense to say that scientists apply mathematical concepts to other mathematical concepts or to physical phenomena with the aim of complying with the regulative and constitutive demands of permanence, generalization, and some others. The (well-known) effectiveness of such application is explained, according to my reading of  Cassirer’s Kantianism, by virtue of employing \emph{a priori} principles of theory development inherently incorporated into the way scientists think, such as the method of mathematical conjecture mentioned in the introduction or any other method of theory construction. 
\section{Cassirer meets Lie group theory}\label{section4}
Let us now mould the previous discussion to fit an explicit form. It would be wise to instantiate the \emph{a priori} principles of theory development, as regards our current stage of knowledge, which allow us to explain the effectiveness of many mathematical theories in its application to physical science. To restrict our analysis to a particular case study, let us consider Cassirer’s best candidate: the mathematical language of \emph{Lie group theory}.\footnote{Relevant definitions of this mathematical language shall be given below.} As argued by Steven French in \citep{french} and \citep[ch.5]{french1}, Lie group theory has been effective in elaborating formulations or reformulations of many physical theories capable of generating new empirical predictions and of improving our understanding about these theories. Granted this, my aim is to identify and characterise certain \emph{a priori}, relativised principles which explain the effectiveness of Lie group theory in its successive application to the physical sciences, and enable the development of certain areas of physical science. As emphasised by \citep{cassirer1922,heis2014b}, these principles must be compatible with both the function theory of concepts and some of the neo-Kantian criteria elucidated above: the regulative ideal of permanence and generality, and the constitutive ideal of explaining the empirical and interpretative sides of physical theories. Considering this condition, we shall see that these principles are represented by group-theoretic properties of certain Lie groups ---\emph{invariance groups} as defined below--- responsible for the effectiveness of applying Lie group theory to physics. 

Let us proceed to complete the following threefold task: firstly, we start by identifying the general mathematical framework in terms of which Lie group theory is framed; secondly, we reveal the structural nature of this mathematical language and its association with the function theory of concepts; and thirdly, we finally proceed to articulate the sense in which certain Lie groups are associated with \emph{a priori}, relativised principles that play both a regulative and constitutive role and explain the effectiveness of applying Lie group theory to physical science.
\subsection{The basics of Lie group theory}
Simply put, a group G consists in a set of (finite or infinite) elements with an operation * defined between two of these elements to form a third element within the group by means of a `multiplication rule’ $x*y=z$, where $x, y, z \in G$. The set and the operation must satisfy the associative property and must have both an identity element and an inverse element. If the group is a differentiable manifold and the operation * is smooth (i.e., it defines a continuous mapping with derivatives of all orders) then it is called a \emph{Lie group}. As one can easily note, this abstract definition does not entail a straightforward physical interpretation. However, this interpretation can be elaborated by means of the following  observation:

There are geometrical objects upon which a Lie group acts. These can be figures composed of spatial points, such as triangles or squares, but they can also be vectors, differential forms, tensors, etc. The action of a Lie group on a set of these objects (e.g., a vector space, an affine space, etc.) is a ‘copy’ of the Lie group defined in that set, in the same way a self-portrait is a copy of the artist represented in a canvas. In the particular case of a Lie group G acting on a vector space V (a case which will be our main focus in this paper), the action of G on V is, in formal terms, a \emph{Lie representation} of G in V: a function that goes from G to the group of automorphisms of V. Those automorphisms are realised as a set of linear transformations or mappings between initial and final points in V. For example, the set of rotations in three-dimensional space are linear transformations induced by the Lie representation of the set of $3\times3$ orthogonal matrices with determinant equal to one, called the \emph{special orthogonal group} SO(3), on $\mathbb{R}^{3}$. 

Thus, in general we have two conceptually different mathematical objects that are equivalent (up to a Lie representation): the abstract Lie groups, on the one hand, and the induced transformations in the mathematical space in which these Lie groups are Lie-represented, on the other. Considering this equivalence relation, \citep{weyl} discovered (through what is known as Weyl's programme) that Lie groups are abstract mathematical objects that have no physical interpretation unless they act upon the state space V of a physical theory T (hereinafter, vector space V of T). As demonstrated by him, the way this interpretation is revealed is through the concept of invariance: the Lie representation of a Lie group G in V takes the form of linear transformations that leave certain sorts of objects invariant, which are interpreted as physical symmetries. In particular, the induced transformations of G can be associated with the dynamical evolution of a physical system described by the laws of T. 

Based on this association, we say that the abstract Lie group G is an \emph{invariance group} of a theory T describing some physical system if the fundamental laws of T are invariant under the induced continuous transformations of G in the vector space V of T (i.e., the Lie representations of G in V). Since the laws of a theory T can be reformulated in this way ---as invariants of a certain invariance group G---, the identification of G (together with the vector space V of T in which G is Lie-represented) is sufficient to determine its dynamical structure, and furthermore, the complete laws of T can be obtained by Lie-representing G in V. 

Once we have introduced the central technical elements of Lie group theory, let us proceed to address our second task.  
\subsection{The structural role of Lie group theory}
Based on the fact that any Lie group element induces a group transformation on a vector space, we can interpret any Lie group as an abstract mathematical object that generates (physical) binary relations on the vector space in which it is Lie group-represented. For example, any simple rotation in the three-dimensional Euclidean space, induced by any element of SO(3) acting on that space, can be interpreted as a binary relation between the initial and the final coordinates of the rotation angle. This alone suffices to convince us that any Lie group is an abstract mathematical object that induces a physical interpretation intrinsically relational or structural when they are Lie group-represented in vector spaces. 

But we can advance this claim as regards the relational structure of Lie groups themselves without being Lie group-represented in vector spaces. According to Arthur Eddington, the Lie group language expresses (in abstract terms) the second-order relationships that hold between physical relations. In his own words, ``Whatever the nature of the entities, the use of group theory allows us to abstract away the ‘pattern’ or structure of relations between them. What the group-structure represents, then, is the ‘pattern of interweaving’ or ‘interrelatedness of relations' \citep[137-140]{eddington}.'' This can be illustrated by appealing to the multiplicative law of SO(3), where any element of this Lie group can be obtained by the composition of two or more elements of the same group, inducing successive and continuous rotations in the three-dimensional Euclidean space. Therefore, group-theoretic structures underlying (or compatible with) our best scientific theories are likely to be relational or structural in this second-order sense. 

This observation suggests that Lie group theory can be interpreted as an appropriate representational framework in terms of which the function theory of concepts can be instantiated. Since this function theory, according to \citep{cassirer1922,heis2014b}, takes the category of structures and relations as more fundamental than the individual concepts themselves, it is important to be explicit with respect to which aspects of scientific knowledge we are making this epistemic commitment. In this way, we conclude that the function theory of concepts can be instantiated appropriately by focusing on those theoretical relations and structures represented by the mathematical theory of Lie group theory. Once we have completed our second task, let us now address our third final task. 
\subsection{The effectiveness of Lie group theory}
Considering that invariance groups can be interpreted as mathematical instantiations of the function theory of concepts, our next step is to articulate the precise way some properties of these groups ultimately explain the effectiveness of applying Lie group theory to physical science. In so doing, we shall identify two \emph{a priori} principles of theory development. 

Firstly, there is a \emph{principle of structural continuity} mathematically realised by the continuous nature of invariance groups (in the inter-theoretic sense), which helps to explain why these mathematical entities necessarily embrace many theoretical domains under an approximately permanent, generalising framework without leaving aside the explanation of the empirical phenomena confined to these domains. Since the necessary commitment to this principle conduces scientists to pursue relativised ideals of permanence and generality whilst accounting for the relevant empirical phenomena, we shall demonstrate that it plays both a regulative and constitutive role. 

Secondly, we shall argue in favour of the claim that group-theoretic conjecture in the mathematical side, interpreted as an \emph{a priori} principle of theory development, necessarily brings about \emph{surplus mathematical structure} that eventually might turn out to be empirically relevant. Similarly, this principle shall be demonstrated to play both a regulative and constitutive role by virtue of the fact that at every step of theory development it necessarily conduces scientists to look for general mathematical frameworks capable of making novel predictions. 

Finally, after addressing some objections raised against the idea that these principles are satisfied by invariance groups, we shall see that the \emph{a prioristic} and necessary natures of these group-theoretic principles shall be the defining elements of my neo-Kantian framework of mathematical applicability applied to the theory of Lie groups. 
\subsubsection{Principle of structural continuity}
Invariance groups are those mathematical entities that are sufficiently general and specific at once. In compliance with category theory (i.e., one of the most general and abstract mathematical frameworks), Lie group theory is sufficiently general in scope and expressive power as it can embrace many mathematical objects, such as rings, fields and vector spaces, in addition to the fact that many physical theories are (or can be) framed in terms of this mathematical language (e.g., classical, quantum and relativistic kinematics, the group-theoretic formulation of general relativity, the standard model of particle physics, gauge theories, etc.). However, in contrast to category theory, it is sufficiently specific and can make a detailed explanation of the empirical phenomena confined to the domains to which it applies (something which shall be  corroborated below following what is known as the Wigner's programme). 

This middle ground between generality and specificity is what makes invariance groups approximately continuous and explanatory, in the sense that they approximately preserve a core set of group-theoretic structures that is not lost in transitions among physical theories and whose main feature is to indispensably explain all the predicted phenomena confined to those theories. This does not mean that the complete set of group-theoretic structures underlying our best theories is immune to future change; it only means that some relevant and indispensable `working parts' of these structures are fully preserved throughout the journey. As discussed by Heinz Post (in the object-oriented context),
\begin{quote}
 ``We may divide a theory `vertically’ into well-confirmed working parts and others. Thus the phlogiston theory `worked’ in that it assigned consistent levels of phlogistication (explaining many features) to chemical substances related in more than one way by reactions. On the other hand it tried to establish a connection between colour and phlogistication, and this part of the theory was not successful even at the time \citep[228]{post}.'' 
\end{quote}
These `working parts' are sometimes (but not necessarily) associated with what is called `working posits', which are theoretical terms (pertaining to physical theories in the context of physics) that are primarily responsible for and indispensable to the explanation and prediction of the relevant phenomena.\footnote{Although working posits are generally introduced and discussed in the context of selective realism, I shall consider their definition in the neo-Kantian context.} More specifically, the associated, preserved working posits are able to explain the alleged successes with respect to the restricted domain in which these theories are relevant. 

In the structural context, this definition suggests that there might be dispensable mathematical structures that are not preserved from one preceding theory to its successor, but there cannot be structural working posits that are lost among theoretical transitions: if structural working posits are lost in the transition, the structure underlying (or compatible) with the successor theories would be incapable of explaining the relevant domain entailed by the preceding theories. A typical example of a structural working posit is the preserved structure of the eikonal particle-like equation among the theoretical transition from Fresnel's theory of light  (for short wave-lengths) to Maxwell's wave electrodynamics (for arbitrary wave-lengths) \citep{worrall}. On the other hand, nontypical examples (where invariance groups are involved) are the metaplectic group-theoretic structure associated with both classical and quantum mechanics \citep{gosson} and the extended Galilean group-theoretic structure associated with both classical and quantum kinematics \citep{green}. 

Considering this continuity property of invariance groups, we can make explicit the first \emph{a priori} principle of theory development, which explains the effectiveness of Lie group theory in its application to physics: 
\begin{quote}
{\bf \emph{Principle of structural continuity}}: other things equal, if the underlying (or compatible) mathematical structure of a successor theory approximately preserves the structural working posits of a predecessor theory, then we necessarily favour ----for the purpose of developing and generalising further science--- that mathematical structure of the successor theory. 
\end{quote}
Let us emphasise that, from a regimented standpoint, this principle is interpreted as a predicate which have non-empty extensions, provided these extensions are associated with mathematical structures underlying (or compatible with) successor theories that approximately preserve the structural working posits of predecessor theories. The claim (already justified above) that invariance groups are part of such extensions conduces us to conclude that the principle of structural continuity, interpreted in the neo-Kantian and transcendental way, is responsible of explaining the effectiveness of Lie group theory in its application to physics. 

Conceived in this way, this principle is the transcendental and structural realisation of Heinz Post's heuristic \emph{general correspondence principle}, according to which ``any acceptable new theory L should account for the success of its predecessor S by `degenerating’ into that theory under those conditions under which S has been well confirmed by tests'' \citep[228]{post}. Similarly, although we should note that our principle possesses an \emph{a prioristic} and necessary (as opposed to heuristic) nature, it retains the spirit of Simon Saunders' structural notion of heuristic plasticity, which is based on the claim that 
\begin{quote}
``What is taken over from preceding theories is not only those laws and experimental facts which are well-confirmed, but also ‘patterns’ and ‘internal connections’, that in this way the successor theory accounts for whatever success its precursor enjoyed.'' \citep[295]{saunders2}. 
\end{quote}
Therefore, the fact that there is at least an \emph{a priori} principle of theory development in the form of a continuity property, which necessarily conduces scientists to look for the underlying invariance groups that preserve certain theoretical and empirical content, is what explains the effectiveness of employing Lie group theory as means for developing and generalising current physical theories in the absence of new empirical data. Let us finally explain why this principle is regarded as both regulative and constitutive.

Note that behind the principle of structural continuity lies the prescription that physical theories should be formulated by certain mathematical structures that are both approximately continuous and explanatory, in the sense that they approximately preserve the structural working posits of physical theories, provided these posits indispensably explain all the predicted phenomena confined to those theories. There are at least two conclusions drawn from this observation. 

Firstly, as long as there are preserved working posits in the form of mathematical structures there is a sense in which some of these structures remain unchanged (i.e., they are permanent) among theoretical transitions. As emphasised above, the structure underlying (or compatible) with the successor theories should be capable of explaining the relevant domain entailed by the preceding theories and this can only be achieved if the structural working posits of the preceding theories are not lost among their transition to successor theories. Furthermore, this transition can be interpreted as a progressive and cumulative generalisation as long as new structural working posits are incorporated into the mathematical structure underlying (or compatible) with the successor theories. Therefore, we can interpret the alleged principle as regulative in the sense that it prescribes how theories should be formulated based on regulative ideals of permanence and generality associated with the mathematical structures underlying (or compatible) with our physical theories.

Secondly, although the principle of structural continuity plays a regulative role through the expression of the above ideals, I would like to convince the reader that it also plays a constitutive role in that the preserved, structural working posits are perpetually self-renewing, with only relative stopping-points in which these posits explain the empirical and interpretative sides of physical theories. In fact, considering that, at each stage of the development of some physical theories, working posits are mathematical structures that are primarily responsible for and indispensable to the explanation and prediction of the relevant phenomena, this principle establishes what is objective according to a certain domain, and the adoption of a new theory in this development becomes rational by virtue of having a broader mathematical structure than the preceding one, making successively better approximations and explanations of the diversity of experience. 
\subsubsection{Principle of surplus structure}
Furthermore, let us briefly note that invariance groups enable us to have another \emph{a priori} strategy for theoretical development at our disposal that contributes to generate new predictions. Considering physical theories capable of being formulated in group-theoretic terms, this powerful strategy is to focus not only on mathematical structures that are empirically relevant by virtue of having counterparts at the empirical level, but also on those group-theoretic structures that are surplus, in the sense that they are elements of the mathematical formalism that do not entail some sort of phenomena in the observable world directly relevant to the theory. More specifically, this principle is expressed as follows: 
\begin{quote}
{\bf \emph{Principle of surplus structure}}: other things equal, if the underlying (or compatible) mathematical structure of a theory contains surplus mathematical structure with respect to the phenomena entailed by that theory, then we necessarily favour ----for the purpose of developing and generalising further science--- that mathematical structure over and above other empirically relevant mathematical structures that do not contain such a surplus structure. 
\end{quote}
As corroborated by many case studies e.g., the famous discovery of the omega baryon by the eightfold way group-theoretic model \citep[223]{buenofrench}, superpositions of mass by the extended Galilean group \citep{green}, etc., there are theories that can be formulated or reformulated in terms of invariance groups that initially contain \emph{surplus structure} that, nevertheless, turn out to be empirically fruitful at a later point in their development \citep{redhead,partialfrench,buenofrench}. More specifically, invariance groups are significant mathematical sources of potential developments by virtue of the fact that their presence as an initially surplus structure in the mathematical domain suggests a way of searching for counterpart relations at the empirical domain, some of which hold between phenomena that are completely undetectable or unknown for physicists of the time. Interpreting this strategy as \emph{a priori} explains why it has been so effective as it is conceived as a necessary way of theorising, in the sense that scientists necessarily look for certain knowledge gaps in the mathematical side that might be filled by the theory, discovering novel empirical phenomena as a consequence. Let us finally explain why this principle is regarded as both regulative and constitutive. 

Firstly, the principle of surplus structure involves the simultaneous satisfaction of two complementary regulative ideals: generality and permanence. From the standpoint of my neo-Kantian framework, the \emph{a priori} preference towards mathematical structures that contain surplus structure is a transcendental criterion based on the regulative ideal of generality i.e., finding a way to expand the predictive scope of our physical theories though the use of mathematical conjecture. This notion of generality is sometimes associated with the notion of predictive strength, namely, the capability of a physical theory to generate more information about the empirical world. 

However, let us note that predictive strength usually competes against the ideal of simplicity, which aims to capture the same predictive content using the minimum amount of mathematical structure. Indeed, adding surplus content to a mathematical structure normally increases the predictive strength of a theory at the cost of its simplicity, while taking away this surplus content increases simplicity at the cost of predictive strength. The unfortunate problem of this competition is that the ideal of simplicity can also be interpreted as a regulative virtue that scientists necessarily use to develop further science. To illustrate this problem, let us appeal to an interesting example critically analised by \citep{acuna}. 

As corroborated by the history of special relativity, there were at least two empirically-equivalent routes to develop this theory, the first of which was a principle theory, elucidated by Albert Einstein, and the second of which was a constructive theory, simultaneously proposed by Henri Lorentz. Since Einstein's theory explained relativistic phenomena in accordance with the relativity postulate and the light postulate, people thought that such a theory was more simple than its alternative, whilst both competitors shared the same predictive strength. Lorentz' theory, by contrast, was able to tell us how to conceive of the reality behind the phenomena (e.g., time dilations and length contractions) at the cost of positing a more robust ontology, such as electrons and the infamous ether. Since Lorentz' theory was eventually ruled out by the emergence of quantum field theory, we can say that the choice of the correct theory was (partly) made on the basis of simplicity in a time when people thought that both theories shared the same predictive strength. From this case study, it follows that the \emph{a priori} criterion that scientists necessarily use to explain the successful development of a physical theory remains underdetermined: we are unable to determine whether such a criterion corresponds to predictive strength or simplicity. 

Under these circumstances, I sustain that we can brake the alleged underdetermination in favour of conceiving predictive strength (as opposed to simplicity) as an \emph{a priori} regulative ideal. In order to do so, we have to demonstrate that the principle of surplus structure is inherently associated with another regulative ideal, which correlates with predictive strength in a virtuous way, whilst competes against simplicity: the ideal of permanence. Considering this additional virtue as it is defined above, one might realise that the principle of surplus structure reinforces the idea that there should be preserved structures that never loose content and expand as it evolves in time. To see this, let us note that retaining surplus content reduces the risk of having to change the entire mathematical structure at the cost of its simplicity, while taking away this surplus structure increases simplicity at the cost of renouncing to preserved mathematical structures. Thus, we conclude that the principle of surplus structure can be regarded as regulative because it involves the simultaneous satisfaction of predictive strength and permanence as \emph{a priori} regulative ideals.  

Secondly, the principle of surplus structure plays a constitutive role in the same way as the principle of structural continuity: in order to satisfy the regulative ideals of predictive strength and permanence, the mathematical structures underlying (or compatible) with any physical theory must 
explain the empirical and interpretative side of this theory.  

Let us now address some critical comments that might be posed against the view that the above continuity principles are satisfied by invariance groups. 
\subsubsection{The regulative and constitutive roles of invariance groups}
Some scholars have supported the claim that invariance groups do not play a constitutive role ---although they accept that these groups actually play a regulative role. Were this claim correct, my conclusion that the above continuity principles are satisfied by invariance groups would be undermined as these groups cannot satisfy something which has been defined as constitutive in the first place. Considering this line of reasoning, let us briefly explain the above claim, as formulated by  \citep{friedman,everett2}, and then elaborate an objection against it.

As argued by \citep{friedman2010a,everett2}, permanence and generality are regulative principles encoded by certain group-theoretic properties of invariance groups. In compliance with the ambitious programme developed by \citep{weyl}, they argue that the primary aim of the application of Lie group theory to physics is to lay the foundations of physical theories driven by a philosophical intuition regarding the timeless and universal nature of the laws of physics. In this way, invariance groups, as defined here, represent symmetries and regularities among the diversity of the phenomena in accordance with the regulative ideal of permanence and generality. 

As regards the constitutive role of invariance groups, however, Everett argues that any element of these group-theoretic structures is not sufficient for capturing the rationality involved in actual practice because it is only capable of providing mathematical form to the theory, i.e., it ``does not tell us to what this mathematical structure corresponds'': it does not explain the empirical and interpretative side of theories \citep[151]{everett2}. Analogously, in his late account of the constitutive \emph{a priori}, Friedman underestimates the constitutive role of invariance groups and provides instead central emphasis to physical ‘coordination principles’ (i.e., conditions of applicability of these invariances) that coordinate the purely abstract, mathematical structures of a theory with its empirical component, the former of which is intended to apply to the latter to furnish ``the necessary framework within which the testing of properly empirical laws is then possible'' \citep[83-79]{friedman2}. For example, as opposed to the \citep{ryckman} reading of \citep{cassirer1923}, \citep{friedman,everett2} do not think that general covariance plays a constitutive role in the explanation and prediction of the phenomena relevant to the relativistic domain. 

Although I agree that invariance groups are not sufficient for establishing a full coordination between the mathematical and empirical domains, I disagree with the claim that they do not play a constitutive role. \emph{Contra} Everett and Friedman, I argue that invariance groups are indispensable not only with respect to the regulative role of demanding permanence and generality, but also with respect to the constitutive role associated with predictive and explanatory power. To defend this argument, we shall present a very important concept associated with the empirical consequences of Lie group theory. 

Let us define the \emph{Lie algebra} of a Lie group G as the tangent space of G at the identity (i.e., a group-theoretic structure which allows one to define the local group structure of G in a vector space). More specifically, the associated real Lie algebra $\mathscr{G}$ of a Lie group G is a real vector space (of the same dimension to that of G) with a bilinear, alternating, antisymmetric product map $[\cdot,\cdot]:\mathscr{G} \times \mathscr{G} \rightarrow \mathscr{G}$, called \emph{Lie bracket}, that satisfies some commutation relations: the \emph{Jacobi identity}.\footnote{The Jacobi identity is $\left[X,\,[Y,Z]\,\right]+\left[Y,\,[Z,X]\,\right]+\left[Z,\,[X,Y]\,\right]=0$; for all $X,Y,Z \in \mathscr{G}$.} As demonstrated by \citep{wigner0,wigner1} (what later became known as the Wigner's programme), the Lie algebra $\mathscr{G}$ associated with the invariance group G of a theory T is relevant to the predictive component of T by virtue of the fact that the Lie representation of $\mathscr{G}$ in the vector space V of T is associated with infinitesimal transformations that represent the physical properties relevant to T. For example, in classical mechanics and QM, the infinitesimal transformations induced by the \emph{generators} of the associated Lie algebras (i.e., the basis operators $T_{A}$ of $\mathscr{G}$, where $A=\{1, 2, ...\}$) are spelled out in terms of physical properties of the system (e.g., momentum, energy, etc.). 

Note that this observation implies that the formulation or reformulation of a physical theory in terms of abstract invariance groups and their associated Lie algebras permits not only to reveal the fundamental mathematical structure of that theory, but also to formally represent its complex connection with the empirical domain. As such, invariance groups play a constitutive role because they permit the empirical side of physical theories through their associated Lie algebras. However, this does not mean that physical principles are not required. These principles could be useful for the purposes of explaining the empirical phenomena, although I believe that, unlike invariance groups, they are dispensable, and therefore, epistemologically non-fundamental (in the Kantian sense). 
\section{Concluding remarks}\label{section5}
Overall, interpreting mathematics and its application to physics via Cassirer’s Kantianism provides a third alternative to the problem of mathematical applicability, at least with regard to physical theories. This is because mathematics and physical theory are not independent elements applied to each other according to certain `correspondence rules’ between our representations and the world, but are different modes of a cognitive process of synthesis that constitutes the physical theory and aims at unity, permanence and generality at each stage of the scientific inquiry. In this sense, there is no mystery in the effectiveness of mathematics in its application to physical science because the explanation of this effectiveness is deeply rooted in the \emph{a prioristic}, constitutive and regulative roles of relativised principles, which establish the conditions of possibility for both mathematics and physics, and for the application of the former to the latter.  
\section*{Acknowledgements}
I am grateful to Ladislav kvasz and Pavel Janda for fruitful feedback. Thanks to Michael Pockley for linguistic advice. The work on this paper was supported by CONAHCYT-Estancias posdoctorales por México 2022 (grant 442599), and by the Formal Epistemology---the Future Synthesis grant, in the framework of the Praemium Academicum programme of the Czech Academy of Sciences.

\end{document}